# Principle and Performance Analysis of the Levenberg–Marquardt Algorithm in WMS Spectral Line Fitting

Yongjie Sun [1,2], Pengpeng Wang [1,*], Tingting Zhang [1], Kun Li [2], Feng Peng [2] and Cunguang Zhu [1,*]

[1] School of Physical Science and Information Technology, Liaocheng University, Liaocheng 252000, China; 202021200721@stu.ujn.edu.cn (Y.S.); 2120110516@stu.lcu.edu.cn (T.Z.)
[2] School of Physical Science and Technology, University of Jinan, Jinan 250022, China; 202021200723@stu.ujn.edu.cn (K.L.); sps_pengf@ujn.edu.cn (F.P.)
* Correspondence: wangpengpeng@lcu.edu.cn (P.W.); zhucunguang@lcu.edu.cn (C.Z.)

**Abstract:** Calibration-free wavelength modulation spectroscopy (WMS) is an efficient technique for trace gas monitoring. It is widely used due to its resistance to light intensity fluctuations, strong suppression of low-frequency noise, fast response time, and excellent environmental adaptability. The calibration-free WMS often employs the Levenberg–Marquardt algorithm for spectral fitting to retrieve gas characteristics. However, to the best of our knowledge, an analysis of the main factors affecting the operational effectiveness of the Levenberg–Marquardt algorithm in calibration-free WMS has merely been reported. In this paper, we have systematically analyzed the Levenberg–Marquardt algorithm's operating mechanism in WMS-2$f$/1$f$. The results show that the number of parameters and the estimation errors of the initial parameters are the main factors limiting the retrieval accuracy of the algorithm, which provides some important guidelines for the subsequent optimization of the spectral fitting scheme.

## 1. Introduction

Wavelength modulation spectroscopy (WMS) is a non-contact technique for trace gas monitoring which is widely used due to its high sensitivity and fast dynamic response [1–5]. In conventional WMS, the measurement signal requires calibration with a gas mixture of known concentrations, which limits the application of WMS technology in harsh environments (high temperature or pressure) or where the relevant gas conditions are poorly known.

Researchers have recently developed numerous calibration-free techniques to address this issue [6–12]. Li et al. created a WMS analytical model that can be simulated using readily available laser parameters and an absorption spectrum [13]. This method can eliminate the majority of calibration factors and infer gas parameters. Rieker et al. utilized the first harmonics to normalize the second harmonics (WMS-2$f$/1$f$) and extended the WMS analysis model to the field of combustion monitoring to retrieve the temperature and concentration of water vapor simultaneously [14]. Our research group has proposed a sensitive second harmonic phase angle measurement method capable of background-free gas monitoring [15]. However, the retrieval accuracy of gas parameters using these techniques generally depends on the comprehensive characterization of the target transition's line shape parameters (such as collisional broadening coefficients) and the laser parameters, which typically require extensive laboratory work. This challenge can be avoided by using the methodology of Christopher et al. [16]. They set the target transition's line shape parameters and the gas concentration as parameters to be estimated and used the Levenberg–Marquardt (LM) algorithm to perform spectral line fitting to retrieve the gas's condition. Due to its lower requirements for line shape and laser parameter characterization, this LM methodology has been extensively used for highly sensitive and robust measurements of gas conditions. Combining the LM algorithm with dual-spectroscopy techniques, Li et al. developed a mid-infrared laser-trace gas sensor capable of online monitoring the concentration of multi-component gases [17]. Zang et al. proposed a method to retrieve high-temperature spectral line parameters using the LM algorithm for least-squares fitting absorption spectra in the combustion flow field [18]. Raza et al. developed a dual-species (CO/NH3) sensor based on the LM methodology for high-temperature measurements [19]. However, an analysis of the main factors affecting the operational effectiveness of the LM algorithm in calibration-free WMS has merely been reported. In this paper, we have systematically analyzed the Levenberg–Marquardt algorithm's operating mechanism in WMS-2*f*/1*f*.

## 2. Methodology
### 2.1. WMS-2f/1f Theoretical Model

The injected current of the laser diode is periodically modulated by a high-frequency sinusoidal signal to produce an instantaneous optical frequency $v(t)$ and light intensity $I_0(t)$:

$$v(t) = v + \Delta v \cos(\omega t) \tag{1}$$

$$I_0(t) = \bar{I}_0 \left[ 1 + i_1 \cos(\omega t + \psi_1) + i_2 \cos(2\omega t + \psi_2) \right] \tag{2}$$

where $v$ is the laser center frequency, $\Delta v$ is the modulation depth, $\bar{I}_0$ is the average light intensity, $i_1$ and $i_2$ are the intensity amplitudes of the first- and second-order (normalized by $\bar{I}_0$), $\psi_1$ is the FM/IM phase shift, and $\psi_2$ is the phase shift between the FM and the nonlinear IM.

According to the Beer–Lambert law, the spectroscopic absorbance can be written as:

$$\tau(v(t)) = \frac{I(t)}{I_0(t)} = \exp[-PCLS(T)\phi_v] = \exp[-A\phi_v] \tag{3}$$

where $I(t)$ and $I_0(t)$ are the transmitted and incident laser intensities, $P$ is the total gas pressure, $S(T)$ is the strength of the absorption line at $T$ temperature, $\Phi_v$ is a line-shape function of the absorption spectrum of the gas, $C$ is the concentration of the gas to be measured, $L$ is the length of the optical range, and $A$ is the integrated absorbance.

The $\tau(v(t))$ can be expanded in a Fourier cosine series, with the $H$ term as follows:

$$H_0(v(t)) = \frac{1}{2\pi} \int_{-\pi}^{\pi} \tau(v(t)) d\theta \tag{4}$$

$$H_k(v(t)) = \frac{1}{\pi} \int_{-\pi}^{\pi} \tau(v(t)) \cos(k\theta) d\theta \tag{5}$$

the $I(t)$ signal is input into the lock-in amplifier (LIA) to extract the $X_{nf}$ and $Y_{nf}$ to determine WMS-2*f*/1*f*, given by Equation (6):

$$\text{WMS-}2f/1f = \sqrt{\left(\frac{X_{2f}}{X_{1f}}\right)^2 + \left(\frac{Y_{2f}}{Y_{1f}}\right)^2}$$

$$= \sqrt{\left(\frac{H_2 + \frac{i_1}{2}(H_1 + H_3)\cos\psi_1 + i_2 H_0 \cos\psi_2}{H_1 + i_1\left(H_0 + \frac{H_2}{2}\right)\cos\psi_1 + \frac{i_2}{2}(H_1 + H_3)\cos\psi_2}\right)^2 + \left(\frac{\frac{i_1}{2}(H_1 - H_3)\sin\psi_1 + i_2 H_0 \sin\psi_2}{i_1\left(H_0 - \frac{H_2}{2}\right)\sin\psi_1 + \frac{i_2}{2}(H_1 - H_3)\sin\psi_2}\right)^2} \quad (6)$$

$$= F(A, m, i_1, i_2, \psi_1, \psi_2)$$

from Equation (6), it is clear that the WMS-2*f*/1*f* spectral model is a function of transition integrated absorbance and laser tuning parameters. For an isolated absorption line, the modulation index *m* is defined by $2\Delta v/\Delta v_c$, where $\Delta v_c$ is the full width at half maximum (FWHM) of the gas absorption line.

*2.2. LM Algorithm Overview*

The LM algorithm is a classical optimization algorithm between the steepest gradient descent method and the Gauss–Newton method. Its fundamental concept is to optimize the nonlinear least squares problems by performing differentiation operations on the artificial initialization parameters [20].

The LM-based spectral fitting technique aims to find the optimal solution *β* to minimize the error function between the simulated and measured spectra. The expression of the error function *z* is described as follows:

$$z = \min_{\beta \in R} \sum_{i=1}^{N} \| F_i(\beta) - y_i \|^2 , \quad (7)$$

where *i* = 1, 2, 3, ..., *N* (given a spectral data set), *F*(*β*) is the simulated WMS-2*f*/1*f* spectrum given by Equation (6), and *y* is the measured spectrum. The LM algorithm updates the parameter *β* with the coefficient matrix ($J^TJ + \mu I$) along the negative gradient direction, and the parameter update equation can be described as:

$$\beta_d^{t+1} = \beta_d^t - \left(J^T J + \mu I\right)^{-1} J^T z \quad (8)$$

where *d* = 1, 2, 3, ..., *D* (estimated parameters), *t* + 1 denotes the *t* + 1th iteration, *I* denotes the unit matrix, $\mu$ is the damping coefficient, and $\mu$ is sensitive to the initial parameter's estimated values. If the initial parameter estimates are appropriately chosen, the damping coefficient $\mu$ will be infinitely small. Equation (8) is transformed into the Gauss–Newton equation,

$$\beta_d^{t+1} = \beta_d^t - \left(J^T J\right)^{-1} J^T z \quad (9)$$

where $J^Tz$ is the gradient, $J^TJ$ approximates the Hessian matrix, and the Hessian is the matrix of its second partial derivatives associated with the estimated parameters *D*.

$$Hessian(z) = \begin{bmatrix} \frac{\partial^2 z}{\partial \beta_1^2} & \frac{\partial^2 z}{\partial \beta_1 \partial \beta_2} & \cdots & \frac{\partial^2 z}{\partial \beta_1 \partial \beta_D} \\ \frac{\partial^2 z}{\partial \beta_2 \partial \beta_1} & \frac{\partial^2 z}{\partial \beta_2^2} & \cdots & \frac{\partial^2 z}{\partial \beta_2 \partial \beta_D} \\ \vdots & \vdots & \ddots & \vdots \\ \frac{\partial^2 z}{\partial \beta_D \partial \beta_1} & \frac{\partial^2 z}{\partial \beta_D \partial \beta_2} & \cdots & \frac{\partial^2 z}{\partial \beta_D^2} \end{bmatrix} \approx \sum_{i=1}^{N} \begin{bmatrix} \frac{\partial F_i}{\partial \beta_1} \frac{\partial F_i}{\partial \beta_1} & \frac{\partial F_i}{\partial \beta_1} \frac{\partial F_i}{\partial \beta_2} & \cdots & \frac{\partial F_i}{\partial \beta_1} \frac{\partial F_i}{\partial \beta_D} \\ \frac{\partial F_i}{\partial \beta_2} \frac{\partial F_i}{\partial \beta_1} & \frac{\partial F_i}{\partial \beta_2} \frac{\partial F_i}{\partial \beta_2} & \cdots & \frac{\partial F_i}{\partial \beta_2} \frac{\partial F_i}{\partial \beta_D} \\ \vdots & \vdots & \ddots & \vdots \\ \frac{\partial F_i}{\partial \beta_D} \frac{\partial F_i}{\partial \beta_1} & \frac{\partial F_i}{\partial \beta_D} \frac{\partial F_i}{\partial \beta_2} & \cdots & \frac{\partial F_i}{\partial \beta_D} \frac{\partial F_i}{\partial \beta_D} \end{bmatrix} \quad (10)$$

If the initial parameter estimates are not appropriately chosen, the damping coefficient $\mu$ will keep increasing. Equation (8) is reduced to the steepest gradient descent method equation,

$$\beta_d^{t+1} = \beta_d^t - (\mu I)^{-1} J^{\mathrm{T}} z \quad (11)$$

where $J \in \mathbb{R}^{N \times D}$ is the Jacobian. This is a matrix containing the first partial derivatives associated with the estimated parameters $D$. $J^{\mathrm{T}}$ is the transposed $J$ matrix, which specifically states that Equation (8) is a function of the Jacobian $J$.

$$J = \begin{bmatrix} \frac{\partial F_1}{\partial \beta_1} & \frac{\partial F_1}{\partial \beta_2} & \cdots & \frac{\partial F_1}{\partial \beta_D} \\ \frac{\partial F_2}{\partial \beta_1} & \frac{\partial F_2}{\partial \beta_2} & \cdots & \frac{\partial F_2}{\partial \beta_D} \\ \vdots & \vdots & \ddots & \vdots \\ \frac{\partial F_N}{\partial \beta_1} & \frac{\partial F_N}{\partial \beta_2} & \cdots & \frac{\partial F_N}{\partial \beta_D} \end{bmatrix} \quad (12)$$

*2.3. The Design Process of the LM-Based Spectral Fitting Technique*

The laser tuning parameter $F(\beta)$ can be characterized in the laboratory according to the procedures in [13], or the laser tuning parameter and the integrated absorbance can be set together as estimated parameters, and the absorption information of gas can be obtained by the LM fit between the measured and simulated spectra. In the optimization process, $\beta$ is defined as $\beta = [m, A, i_1, i_2, \psi_1, \psi_2]$:

Step 1: Initial weights. Let the number of iterations $t = 0$, the adjustment of the appropriate coefficient, $\mu$, initial parameter, $\beta$, and the error tolerance, $\varepsilon$, be set;
Step 2: The Jacobian matrix $J$ is calculated according to the current parameters;
Step 3: During iterative optimization, the parameter $\beta^{t+1}$ is updated by Equation (8);
Step 4: Based on the updated parameters, the error between the simulated and measured spectra is recalculated. If $z > \varepsilon$, adjust the damping factor $\mu$ and return to Step 2. Otherwise, the iteration is terminated, and the results are output.

The flow chart of LM-based spectral fitting for extracting gas absorption information is shown in Figure 1. Once the fitting routine has converged, the best-fit integrated absorbance $A$ is used to calculate the gas concentration:

$$C_{\mathrm{Sim}} = \frac{A}{PS(T)L} \quad (13)$$

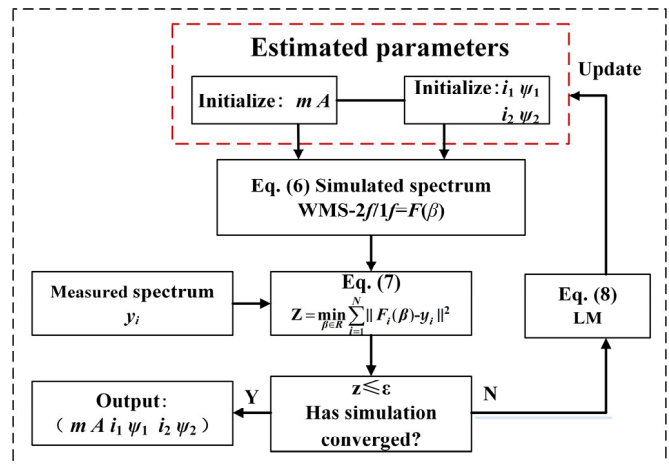

**Figure 1.** Flow chart of LM-based spectral fitting.

## 3. Simulation Results and Analysis

This section analyzes the main factors affecting the operational effectiveness of the LM algorithm in the WMS-2$f$/1$f$ system using the MATLAB R2019b simulation software. In this case, the computer has the following specifications: CPU: i5-10400F; RAM: 8 GB.

The P (13) spectral line of acetylene gas at 1532.83 nm was selected as the simulation target spectral line and periodically scanned to collect a set of reference data, recorded as $y$. The main parameters used in the simulation are listed in Table 1.

**Table 1.** List of model parameters.

| Parameters | Accurate Value |
|---|---|
| $C$ [ppmv] | 296 |
| $i_1$ | 0.20 |
| $i_2$ | $3 \times 10^{-3}$ |
| $\psi_1[\pi]$ | 1.60 |
| $\psi_2[\pi]$ | 1.50 |
| $m$[cm$^{-1}$] | 1.43 |
| $\Delta v_c/2$[cm$^{-1}$/atm] | 0.0777 |

In order to better analyze and evaluate the main factors affecting the operational effectiveness of the LM algorithm in the WMS-2$f$/1$f$ system, the relative error of concentration $C_\delta$, the fitting degree $H$, and convergence time are introduced. The calculation equations are shown in Equations (14) and (15), respectively. It can be seen from the equation that the smaller the relative error of concentration and the fitting degree, the closer the fitted value is to the accurate value, and the better the operational effectiveness of the LM algorithm.

$$H = \sum_{i=1}^{N}(F_i - y_i)^2 \qquad (14)$$

$$C_\delta = \frac{|C_{\text{Sim}} - C|}{C} \times 100\% \qquad (15)$$

### 3.1. Effect of the Number of Estimated Parameters

The effect of the number of estimated parameters in the WMS-2$f$/1$f$ model on the operational effectiveness of the LM algorithm for concentration retrieval is analyzed. Figure 2 shows that when the number of estimated parameters in the WMS-2$f$/1$f$ model increases from two to six (the integrated absorbance *A* is fixed as the estimated parameter), the relative error curve of concentration increases nonlinearly. The specific comparison results are shown in Table 2.

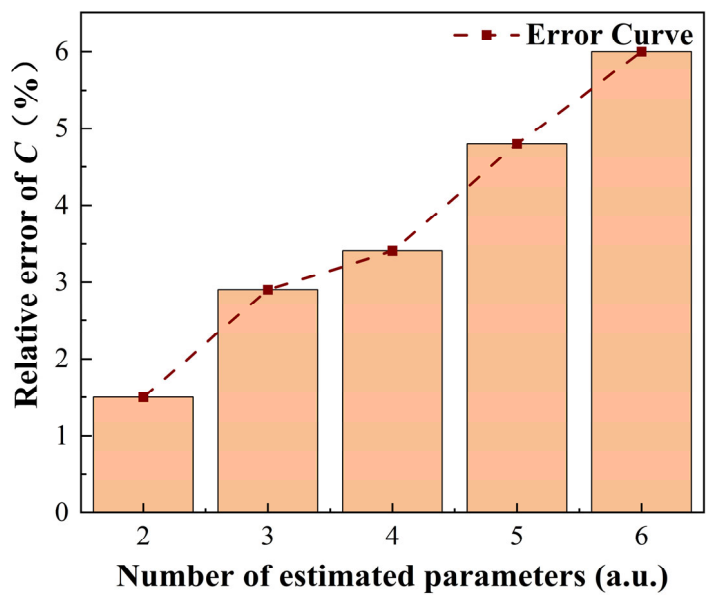

**Figure 2.** The relative error of concentration under different numbers of parameters in the model.

According to Table 2, for two estimated parameters (other parameters not listed are accurate values), $C_\delta$ is 1.5%, $H$ is $3.2 \times 10^{-8}$, and the convergence time is 104 s. With the increase in the number of estimated parameters, the retrieval results in Table 2 become worse. In the case of six estimated parameters, $C_\delta$ is 6.0%, $H$ is $3.0 \times 10^{-3}$, and the convergence time is 236 s. In summary, there is a negative correlation between the operational performance of the LM algorithm and the number of estimated parameters in the WMS-2*f*/1*f* model.

**Table 2.** Comparison results under a different number of estimated parameters.

| Estimated Parameters | Relative Error of Concentration $C_\delta$ (%) | Fitting Degree $H$ | Convergence Time (s) |
|---|---|---|---|
| 2 [$m$, $A$] | 1.5 | $3.2 \times 10^{-8}$ | 104 |
| 3 [$m$, $i_1$, $A$] | 2.9 | $8.7 \times 10^{-7}$ | 140 |
| 4 [$m$, $i_1$, $i_2$, $A$] | 3.4 | $5.4 \times 10^{-6}$ | 173 |
| 5 [$m$, $i_1$, $i_2$, $\psi_1$, $A$] | 4.8 | $1.9 \times 10^{-5}$ | 201 |
| 6 [$m$, $i_1$, $i_2$, $\psi_1$, $\psi_2$, $A$] | 6.0 | $3.0 \times 10^{-3}$ | 236 |

In Section 3.1, we conclude that the accuracy of the retrieval of gas concentration decreases as the number of estimated parameters increasing. We further analyze the reasons for this phenomenon. As mentioned in Section 2.2, the Jacobian *J* is the matrix that is the first partial derivative of the estimated parameters. As the number of estimated parameters used for the calculation increases, the size of the *J* matrix will increase. Additionally, the primary parameter update equation, Equation (8), in the LM algorithm is a function of the Jacobian *J*. The variation in the matrix size of *J* directly leads to an increase in its computational dimensionality, which ultimately affects the operational performance of the LM algorithm for concentration monitoring.

*3.2. Effect of Estimation Errors of the Initial Parameters*

The operational effectiveness of the LM algorithm for concentration retrieval under different estimation errors of the initial parameters (absolute error between the initial parameter, the estimated value, and the accurate value) is analyzed. Figure 3 shows that the relative error curve of concentration increases with the estimation errors of the initial parameters.

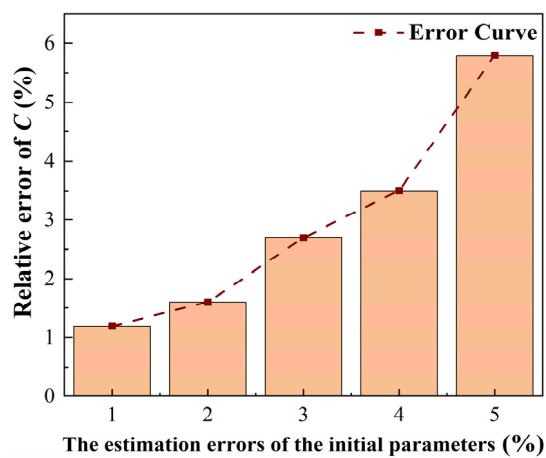

**Figure 3.** The relative error of concentration under the different estimation errors of the initial parameters.

According to Table 3, we analyzed the changes in retrieval accuracy during similar convergence times, using the relative error of concentration and fitting degree as evaluation indicators. In the case of 1% estimation errors, $C_\delta$ is 1.2%, and $H$ is $3.6 \times 10^{-12}$. The overall operating performance of the LM algorithm is good. In the case of 5% estimation errors, $C_\delta$ is 5.8%, and $H$ is $1.4 \times 10^{-9}$. Compared with the 1% estimation error case, $C_\delta$ increases about five times, and $H$ decreases by three orders of magnitude. The overall operating performance of the LM algorithm is poor. In short, a negative correlation exists between the operational performance of the LM algorithm and the estimation errors of the initial parameters.

**Table 3.** Comparison results under the different estimation errors of the initial parameters.

| Estimation Errors of the Initial Parameters (%) | Relative Error of Concentration $C_\delta$ (%) | Fitting Degree $H$ | Convergence Time (s) |
| --- | --- | --- | --- |
| 1 | 1.2 | $3.6 \times 10^{-12}$ | 243 |
| 2 | 1.6 | $1.6 \times 10^{-11}$ | 241 |
| 3 | 2.7 | $4.2 \times 10^{-11}$ | 239 |
| 4 | 3.5 | $8.2 \times 10^{-11}$ | 238 |
| 5 | 5.8 | $1.4 \times 10^{-9}$ | 244 |

Section 3.2 concludes that the accuracy of gas concentration and effectiveness decreases as the estimation errors of the initial parameters increasing. We analyze the reasons for this situation. The LM algorithm is sensitive to the initial parameter estimates used for selection. When the estimation errors of the initial parameters are small, the initial parameter estimated values are sufficiently close to the accurate values. As mentioned in Section 2.2, the damping coefficient $\mu$ will be infinitely small, and Equation (8) is transformed into the Gauss–Newton equation (9). The Gauss–Newton algorithm with second-order convergence performance dominates the LM algorithm. The Taylor expansion of the function $F$ ignores the higher-order derivatives of the estimation errors, making a linear approximation in the neighborhood of the initial parameter estimated values, and converting the nonlinear least squares problem into a linear least squares problem for the optimal solution *β*. The comparison results in Table 3 (estimation errors ≤ 4%) show that the overall operating performance of the LM algorithm is good. In the opposite case, when the estimation errors of the initial parameters are large, the initial estimated parameter values are far from the accurate values. The damping coefficient $\mu$ will keep increasing, so Equation (8) is reduced to the steepest gradient descent method, Equation (11). The steepest gradient descent algorithm with first-order convergence performance dominates the LM algorithm. The comparison results in Table 3 show that the LM algorithm performs poorly in the operation of concentration monitoring.

From the analysis in Section 3, we know that both the number of parameters and the estimation errors of the initial parameters are negatively related to the operational performance of the LM algorithm. For cases where the accuracy of the gas concentration retrieval decreases with the increase in the estimated parameters used for calculations, we generally pre-characterize the laser tuning parameters in all estimated parameters to address it, which typically requires extensive laboratory work. This workload problem can be well avoided if the reference signal without gas absorption is employed for normalization. During normalization, their common variations, such as the laser's intensity fluctuation and phase shift, are eliminated, and the linearity of the spectral model is effectively improved. For example, the authors of [21] described a logarithmically transformed two-beam scheme in which *F* is a function of the absorption line shape. It is not necessary to consider the laser tuning parameters.

For this situation, the LM algorithm is sensitive to the choice of the initial parameter estimates and generally requires continuous debugging to determine the optimal parameters. However, this method is greatly affected by human factors and has lower efficiency. Swarm intelligence algorithms such as the particle swarm optimization algorithm, the artificial fish swarm algorithm, and the ant colony optimization algorithm, which have become more popular non-linear optimization algorithms in recent years, have been widely recognized for their effectiveness and accuracy. Compared with classical LM algorithms, swarm intelligence algorithms can effectively realize the automatic selection of optimal parameters through information interaction between populations. Therefore, the spectral fitting technique based on the swarm intelligence algorithm may help to improve the convergence efficiency and reduce the estimation errors of the initial parameters.

## 4. Conclusions

This paper analyzes the operational principle of the LM algorithm for spectral line fitting in WMS-2*f*/1*f* and further discusses the main factors affecting the retrieval effect of the LM algorithm. The results show that both the number of parameters and the estimation errors of the initial parameters are negatively related to the operational performance of the LM algorithm. In addition, this paper focuses on analyzing the reasons for this phenomenon based on the theoretical formulation derivation and provides targeted solutions for the issues presented in the LM algorithm. For example, pre-characterization processing and reference signal normalization are employed to reduce the number of estimated parameters. A swarm intelligence algorithm is used to reduce the estimation errors effectively.

Finally, as there are more estimated parameters in the fitting routine, the LM algorithm will perform worse. Currently, the characterization workload will be reduced. On the contrary, there are fewer estimated parameters in the fitting routine, which means that the characterization workload will increase, and the overall performance of the LM algorithm will improve. Therefore, the number of estimated parameters and the estimation errors of the initial parameters need to compromise the operational effectiveness of this algorithm and the characterization workload. This paper provides some important guidelines for the subsequent improvement and optimization of the spectral fitting scheme.

preparation, Y.S.; writing—review and editing, C.Z., Y.S., P.W., and F.P.; funding acquisition, C.Z. and F.P. All authors have read and agreed to the published version of the manuscript.

**Funding:** This research was funded by the National Natural Science Foundation of China (No. 61705080), the Promotive Research Fund for Excellent Young and Middle-aged Scientists of Shandong Province (Nos. ZR2016FB17 and BS2015DX005), the Youth Innovation Technology Support Program of Shandong Province Colleges and Universities under Grant (No. 2019KJJ011), the Scientific Research Foundation of Liaocheng University (No. 318012101), and the Startup Foundation for Advanced Talents of Liaocheng University (Nos. 318052156 and 318052157).

**Institutional Review Board Statement:** Not applicable.

**Informed Consent Statement:** Not applicable.

**Data Availability Statement:** The data underlying the results presented in this paper are not publicly available at this time but may be obtained from the authors upon reasonable request.

**Conflicts of Interest:** The authors declare no conflict of interest.